\documentclass[twocolumn]{jpsj2} 
\title{Effect of Uniaxial Stress for Pressure-Induced Superconductor SrFe$_2$As$_2$}

\author{Hisashi \textsc{Kotegawa}$^{1,4}$\thanks{E-mail address: kotegawa@crystal.kobe-u.ac.jp}, Takayuki \textsc{Kawazoe}$^{1}$, Hitoshi \textsc{Sugawara}$^{2}$, Keizo \textsc{Murata}$^{3}$, and Hideki \textsc{Tou}$^{1,4}$}

\inst{$^{1}$Department of Physics, Kobe University, Kobe 657-8501\\
$^{2}$Faculty of Integrated Arts and Science, Tokushima University, Tokushima 770-8502\\
$^{3}$Division of Molecular Materials Science, Graduate School of Science, Osaka City University, Osaka 558-8585, Japan\\
$^{4}$JST, Transformative Research - Project on Iron Pnictides (TRIP), Chiyoda, Tokyo 102-0075\\
}

\abst{
We report that the pressure-temperature phase diagram of single-crystalline SrFe$_2$As$_2$ is easily affected by the hydrostaticity of a pressure-transmitting medium. For all of the three mediums we used, superconductivity with zero resistance appears, accompanied by the suppression of an antiferromagnetic (orthorhombic) phase, but the critical pressure $P_c$ was found to depend on the type of medium. $P_c$ was estimated to be 4.4 GPa under almost hydrostatic condition, but it decreased to $3.4-3.7$ GPa with the use of the medium already solidified at room temperature.
The uniaxial stress along the $c$-axis is suggested to aid in the suppression of the antiferromagnetic (orthorhombic) phase. The pressure effect of BaFe$_2$As$_2$ is also reported.
}

\kword{SrFe$_2$As$_2$, superconductivity, pressure, single crystal}

\begin{document}
\maketitle

Pressure application  has been an effective tool in research on Fe-based superconductors.
The LaFeAs(O$_{1-x}$F$_x$) system shows a significant increase in the superconducting transition temperature $T_c$.\cite{Takahashi}
In FeSe, likewise,  $T_c$ increases to more than three times its original value with pressure application.\cite{Mizuguchi,Margadonna,Medvedev,Masaki}
On the other hand, some Fe-based mother materials, which are stoichiometric and non-superconducting, have been reported to show pressure-induced superconductivity.
SrFe$_2$As$_2$ is one such pressure-induced superconductor.\cite{Alireza,Igawa,Kotegawa}
In our previous study,\cite{Kotegawa} SrFe$_2$As$_2$ shows a zero-resistance state below 34 K at a pressure above 3.5 GPa, accompanied by the suppression of its antiferromagnetic (AFM) phase.
The critical pressure of the boundary between the AFM phase and the paramagnetic (PM) phase was estimated to be $P_c = 3.6-3.7$ GPa.
However, the pressure ranges where superconductivity appears differed among research groups.
Alireza {\it et al.} reported that a diamagnetic signal appears above 3 GPa using a miniature diamond anvil cell.\cite{Alireza}
They used single-crystalline samples and Daphne7373 as a pressure-transmitting medium.
Igawa {\it et al.} reported the zero-resistance state at 8 GPa using polycrystalline samples, and Fluorinert (FC-77:FC-70 = 1:1) was used as the pressure-transmitting medium.\cite{Igawa}
In our experiment, we used single-crystalline samples, Daphne7373, and an indenter cell.\cite{indenter}
It has been reported that the compressibility of SeFe$_2$As$_2$ is anisotropic.\cite{Kumar}
Thus, it is expected that the system is sensitive to uniaxial stress of pressure.
Measurements under high pressure are important in research on Fe-based superconductors, but we need to understand what effect is induced by the uniaxial stress in these two-dimensional systems.

In this study, we have investigated the change in the pressure-temperature phase diagram of SrFe$_2$As$_2$ for different pressure-transmitting mediums.
In the case of an almost hydrostatic pressure, it was found that $P_c$ increases to 4.4 GPa and the zero-resistance state appears above $\sim4$ GPa.
If we use a pressure-transmitting medium that solidifies at lower pressures, $P_c$ tends to decrease.
From the direction of the sample in the pressure cell, the stress along the $c$-axis is suggested to promote the suppression of the AFM (orthorhombic) phase.

Single-crystalline samples of SrFe$_2$As$_2$ (BaFe$_2$As$_2$) were prepared by the Sn flux (FeAs flux) method as reported in refs. 11 and 12.
Electrical resistivity measurement at high pressures was carried out using an indenter cell.\cite{indenter}
Resistivity ($\rho$) was measured by a four-probe method, while introducing a flow of current along the $ab$-plane.
Daphne7373, Daphne7474, and Stycast1266 were used as pressure-transmitting mediums.
Daphne7373 has been reported to solidify at $P_{sol}=2.2$ GPa at room temperature, and Stycast1266 was used after it had already polymerized.
Daphne7474 is reported to solidify at $P_{sol}=3.7$ GPa at room temperature and $P_{sol}=6.7$ GPa at 100 ${\rm ^{\circ}C}$.\cite{Murata}
Applied pressure was estimated from the $T_{c}$ of a lead manometer.

\begin{figure}[htb]
\centering
\includegraphics[width=0.8\linewidth]{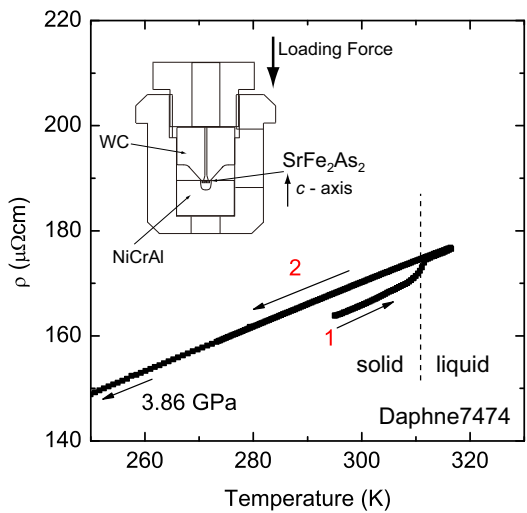}
\caption[]{
(color online) Temperature dependence of $\rho$ for Daphne7474 at approximately room temperature.
After clamping the nut of the pressure cell, the uniaxial stress along the $c$-axis is induced in the sample, but it is relaxed by warming up and liquefying the medium.
The schematic picture of the indenter cell is shown. The $c$-axis of the single-crystalline sample faces in the same direction as the loading force, inducing the uniaxial stress along the $c$-axis above $P_{sol}$.
}
\end{figure}

Figure 1 shows the temperature dependence of $\rho$ at approximately room temperature in the case of Daphne7474.
A schematic picture of the indenter pressure cell is also shown.\cite{indenter}
In the pressure cell, a single-crystalline sample with a thin plate like shape, was placed with its $c$-axis oriented in the same direction as the loading force.
The loading force reduces the sample space, which is a hole in NiCrAl, in the same direction as the $c$-axis.
It is considered that the stress along the $c$-axis is induced above $P_{sol}$.

In the case of Daphne7474, we tried to realize an almost hydrostatic pressure by utilizing the increase in $P_{sol}$ at higher temperatures.
At high pressures, Daphne7474 is solidified after clamping a nut onto the cell to maintain pressure.
First, the cell is warmed up (process 1 in the figure), and then $\rho$ shows a rapid increase at approximately 310 K.
This temperature corresponds to the transition temperature of Daphne7474 from solid to liquid.
The stress along the $c$-axis is expected to exist below 310 K, but it can be relaxed by the liquefaction of Daphne7474.
After that, the cell was cooled down (process 2), and the pressure was estimated at low temperatures.
This process was performed at each pressure up to the highest pressure in this experiment.

\begin{figure}[htb]
\centering
\includegraphics[width=0.85\linewidth]{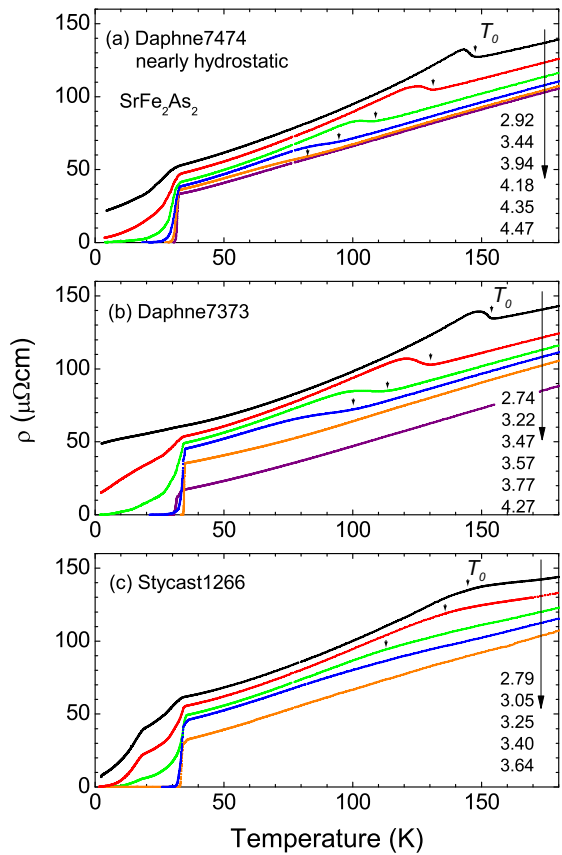}
\caption[]{
(color online) Temperature dependences of $\rho$ in SrFe$_2$As$_2$ for (a) Daphne7474 ($P_{sol}=3.7$ GPa at room temperature and $P_{sol}=6.7$ GPa at 100 ${\rm ^{\circ}C}$), (b) Daphne7373 ($P_{sol}=2.2$ GPa), and (c) Stycast1266 (already polymerized).  $T_0$ are indicated by arrows. In all cases, the anomaly at $T_0$ disappears at high pressures and superconductivity appears accompanied by the disappearance of the anomaly at $T_0$.
}
\end{figure}

Figure 2 shows the temperature dependences of $\rho$ at high pressures using (a) Daphne7474, (b) Daphne7373, and (c) Stycast1266.
$\rho$ obtained using Daphne7373 is the same as that in ref. 8.
In the cases of Daphne7474 and Daphne7373, $\rho$ shows a small jump at $T_0$, which indicates the transition temperature from the PM (tetragonal) phase to the AFM (orthorhombic) phase.
In the case of Daphne7474, the anomaly at $T_0$ survives even at 4.35 GPa, but we cannot find the corresponding anomaly at 4.47 GPa.
The critical pressure $P_c$ is estimated to be approximately $4.4$ GPa.
The onset of superconductivity was observed at 2.92 GPa, but the zero-resistance state is realized above $\sim4$ GPa.
Here, we define the temperature of zero resistance as $T_c$, because resistivity is sensitive to partial superconductivity, and it is difficult to discuss the phase diagram using the onset temperature.
$T_c$ increases with increasing pressure toward 4.47 GPa, reaching 30.1 K.
In this experiment, unfortunately, applying further pressure was difficult.
The $P_c$ of 4.4 GPa contradicts the previous result using Daphne7373.\cite{Kotegawa}
In the case of Daphne7373, the anomaly at $T_0$ disappears above $P_c = 3.6-3.7$ GPa, and the zero-resistance state was observed above $\sim3.5$ GPa.
We notice that $\rho$ for Daphne7373 is lower than that for Daphne7474 in the whole temperature range at high pressures, particularly above 4 GPa.
Figure 1 shows that $\rho$ decreases under the application of uniaxial stress along the $c$-axis, although the reason for this is unclear.
When the medium is solidified, this tendency becomes marked at high pressures where strong uniaxial stress is likely to be induced.
The difference in the absolute value of $\rho$ between Daphne7373 and Daphne7474 at high pressures indicates that the uniaxial stress is stronger in Daphne7373 than in Daphne7474.
From the difference in $P_c$ between these mediums, the uniaxial stress along the $c$-axis is suggested to make the AFM (orthorhombic) phase unstable.

We used Stycast1266 in order to apply a stronger uniaxial stress to the sample.
It has already been confirmed that applying pressure using Stycast1266 induces a uniaxial stress stronger than that in the case of Daphne7373.\cite{Hidaka}
In ref.~14, Hidaka {\it et al.} investigated the effect of uniaxial stress in PrFe$_4$P$_{12}$ under high pressures, where the structural phase transition from the cubic phase to the tetragonal phase occurs at the metal-insulator transition temperature.
In the low-temperature tetragonal (insulator) phase, the $c$-axis length decreases compared with that in the high-temperature cubic (metallic) phase.\cite{Kawana}
Thus, the tetragonal (insulator) phase is considered to become stable under the stress along the $c$-axis induced by the use of Stycast1266.
They observed that the tetragonal (insulator) phase appears at lower pressures in Stycast1266 than in Daphne7373.\cite{Hidaka}

In addition to the strong uniaxial stress, a large pressure distribution is expected to be induced by Stycast1266.
Therefore, it is difficult to evaluate pressure accurately under such a condition; however, we estimated pressure tentatively from the lead manometer as well as in other transmitting mediums.
In the case of Stycast1266, $\rho$ shows no increase at $T_0$, probably owing to the randomness caused by the pressure distribution or the strong stress.
We determined the peak of $-d^2\rho/dT^2$ as $T_0$, although it has a large error bar.
The anomaly at $T_0$ smears at approximately 3.4 GPa, and $T_0$ is difficult to determine.
The zero-resistance state appears above 3.25 GPa, and $T_c$ is maximum at $3.5-3.6$ GPa; the maximum $T_c$ is realized at 3.77 GPa in the case of Daphne7373.
$P_c$ is estimated to be approximately $3.4-3.5$ GPa from these results.

\begin{figure}[htb]
\centering
\includegraphics[width=0.9\linewidth]{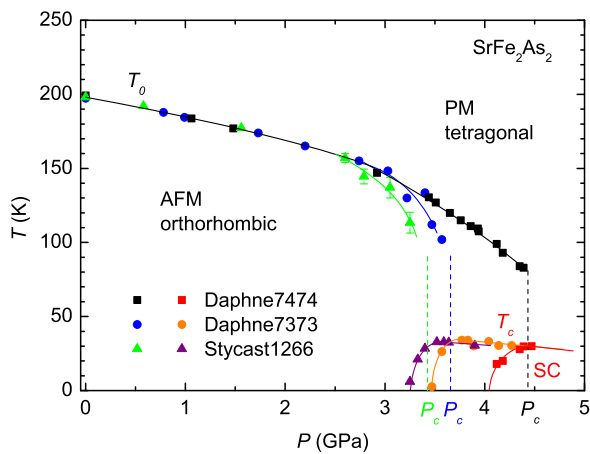}
\caption[]{
(color online) Pressure-temperature phase diagram of SrFe$_2$As$_2$. $P_c$ is estimated to be 4.4 GPa for Daphne7474, $3.6-3.7$ GPa for Daphne7373, and $3.4-3.5$ GPa for Stycast1266. The difference in the pressure dependences of $T_0$ appears above $\sim3$ GPa. In all cases, $T_0$ seems to disappear suddenly from $80-100$ K.
}
\end{figure}

Figure 3 shows pressure-temperature phase diagrams where the results of the three pressure-transmitting mediums are summarized.
The pressure dependences of $T_0$ are consistent with each other among the three mediums below $\sim3$ GPa, but they show differences above $\sim3$ GPa.
$P_c$ is estimated to be 4.4 GPa for Daphne7474, $3.6-3.7$ GPa for Daphne7373, and $3.4-3.5$ GPa for Stycast1266.
The difference in $P_c$ between Daphne7373 and Stycast1266 is small.
This suggests that the AFM (orthorhombic) phase is unstable under uniaxial stress above $\sim3$ GPa.

Such an effect of uniaxial stress can be understood when we consider the structural change between the tetragonal phase and the orthorhombic phase.
Tegel {\it et al.} reported the temperature dependence of the lattice constant of SrFe$_2$As$_2$ at ambient pressure.\cite{Tegel}
The length of the $c$-axis increases when the temperature is decreased to slightly below $T_0$, that is, the length of the $c$-axis in the orthorhombic phase is larger than that in the tetragonal phase in the vicinity of $T_0$.
This implies that the tetragonal phase becomes stable under uniaxial stress along the $c$-axis, which explains the effect of uniaxial stress on $P_c$.
In this context, on the other hand, the stress along the [1 1 0] direction in the tetragonal symmetry is expected to raise $P_c$.
In the polycrystalline sample, the zero-resistance state appears at 8 GPa,\cite{Igawa} implying that a part of the sample has a $P_c$ of approximately 8 GPa.

\begin{figure}[htb]
\centering
\includegraphics[width=0.85\linewidth]{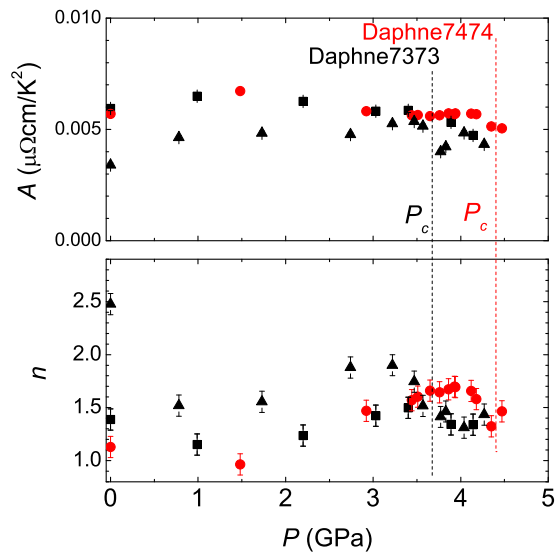}
\caption[]{
(color online) Pressure dependences of the exponent $n$ and the coefficient $A$ in $\rho(T)$ for Daphne7373 (squares and triangles for two different samples) and Daphne7474 (circles). There is no characteristic behavior at approximately $P_c$.
}
\end{figure}

We discuss the behavior at approximately $P_c$ by considering the power $n$ and the coefficient $A$ in $\rho(T)$.
$n$ was estimated using $\rho(T) = \rho_0 + A'T^n$, and $A$ was tentatively estimated using $\rho(T) = \rho_0 + AT^2$ ($n=2$ is fixed), where $\rho_0$ is the residual resistivity.
In both cases, data between 35 and 60 K were used for the fitting to avoid superconductivity and the increase in $\rho$ below $T_0$.
In some heavy-fermion systems with a quantum critical point, the divergence of $A$ and a rapid decrease in $n$ are observed in the vicinity of the critical point.\cite{Demur}
These are typical behavior at the quantum critical point, and spin fluctuations at the quantum critical point are considered to induce the non-Fermi liquid behavior of $n<2$ and to contribute to the occurrence of superconductivity.
Figure 4 shows the pressure dependences of $n$ and $A$ for Daphne7373 and Daphne7474.
The pressure dependence of $A$ is weak, and there is no characteristic behavior at approximately $P_c$.
For both mediums, $n$ exhibits small increases below each $P_c$ owing to the influence of the increase in $\rho$ below $T_0$.
Above $P_c$, $n$ was approximately $1.4$ for both cases.
SrFe$_2$As$_2$ does not exhibit behavior similar to the typical behavior at the quantum critical point seen in some heavy-fermion systems.
It is not clear whether $A$, which was determined at relatively high temperatures of $35-60$ K, is proportional to the square of the density of states at the Fermi level in SrFe$_2$As$_2$, but the absence of the quantum critical behavior is considered to originate from the obvious 1st-order phase transition between the PM phase and the AFM phase.
In the resistivity measurements, $T_0$ disappeared suddenly after it reached $80-100$ K in all cases for the different pressure-transmitting mediums.

\begin{figure}[htb]
\centering
\includegraphics[width=0.9\linewidth]{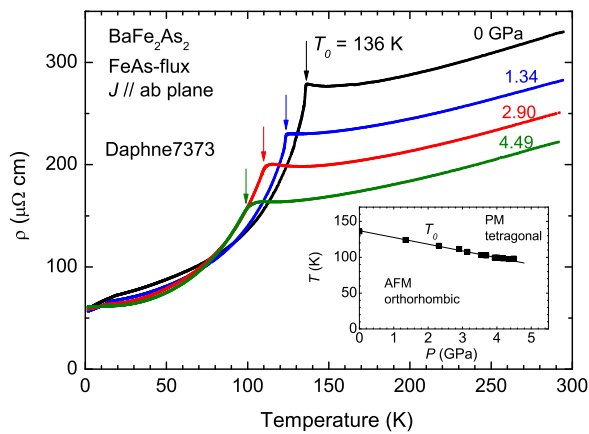}
\caption[]{
(color online) Temperature dependence of $\rho$ for BaFe$_2$As$_2$ grown by FeAs - flux method. The inset shows the pressure dependence of $T_0$. The AFM (orthorhombic) phase survives robustly even at 4.5 GPa, and there is no signature of superconductivity.
}
\end{figure}

Figure 5 shows the temperature dependence of $\rho$ in BaFe$_2$As$_2$ at high pressures up to 4.5 GPa.
This material has also been reported to show superconductivity under pressure by some groups.\cite{Alireza,Mani}
Alireza {\it et al.} reported superconductivity at 29 K above $\sim3$ GPa,\cite{Alireza} while Mani {\it et al.} reported superconductivity at 35 K above $\sim1.5$ GPa.\cite{Mani}
However, no zero-resistance state was observed at pressures of up to 13 GPa in polycrystalline samples.\cite{Fukazawa}
In our experiment, there is no signature of superconductivity up to 4.5 GPa.
The kink at $T_0$ remains distinct even at 4.5 GPa, and $T_0$ shows only a linear decrease, as shown in the inset.
In analogy with the results in SrFe$_2$As$_2$, we conjuncture that the strength of uniaxial stress induces the inconsistency between the experiments.
Actually Mani {\it et al.} used steatite as a pressure-transmitting medium, which is expected to induce a strong uniaxial stress, although we were unable to comprehend which direction the sample faces.
In our experiment using Daphne7373, the uniaxial stress along the $c$-axis exists above approximately $P_{sol}=2.2$ GPa, but $T_0$ decreases just linearly.
A strong uniaxial stress is likely to be needed to suppress the AFM (orthorhombic) phase in this pressure region.

In summary, we have investigated the change in the pressure-temperature phase diagram of SrFe$_2$As$_2$ for different pressure-transmitting mediums.
The critical pressure $P_c$, which is the boundary between the AFM (orthorhombic) phase and the PM (tetragonal) phase at low temperatures, strongly depends on the uniaxial stress along the $c$-axis.
Although $P_c$ was 4.4 GPa in the almost hydrostatic situation using Daphne7474, it was estimated to be $3.6-3.7$ GPa for Daphne7373, and $3.4-3.5$ for Stycast1266.
The difference in $P_c$ between Daphne7373 and Stycast1266 is small, and their pressure dependences of $T_0$ are consistent with each other up to $\sim3$ GPa, indicating that the phase transition at $T_0$ rapidly becomes sensitive to the uniaxial stress above $\sim3$ GPa.
In contrast, if we apply the uniaxial stress along the [1 1 0] direction of the tetragonal symmetry, the AFM (orthorhombic) phase is expected to become stable from its structural character.
We suggest that the sensitivity to uniaxial stress is the main reason why some groups have reported different phase diagrams for SrFe$_2$As$_2$ and as well as BaFe$_2$As$_2$.

We acknowledge T. C. Kobayashi for giving us useful information on Daphne7474.
This work has been partly supported by Grants-in-Aid for Scientific Research (Nos. 19105006, 19051014, 19340103, 19014018, 20740197, and 20102005) from the Ministry of Education, Culture, Sports, Science, and Technology (MEXT) of Japan.

\end{document}